\documentclass[preprint,preprintnumbers,prd,amsmath,amssymb,nofootinbib,tightenlines]{revtex4}
\usepackage{graphicx}% Include figure files
\usepackage{dcolumn}% Align table columns on decimal point
\usepackage{bm}% bold math

\begin{document}
%\preprint{UK/TP 11-xx}

\title{\boldmath%
Superluminal Neutrinos without Revolution
\unboldmath}

\author{Susan Gardner}

\affiliation{
Department of Physics and Astronomy, University of Kentucky, 
Lexington, KY 40506-0055
}

%\date{\today}

\begin{abstract}
The velocity anomaly recently reported by the OPERA collaboration 
appears strikingly at odds with the theory of special relativity. 
I offer a reinterpretation which removes this conflict,  to wit
that neutrinos yield a truer measurement of Einstein's limiting speed, 
and that light and indeed all other matter are retarded by additional 
interactions with the dark universe. 
I discuss existing experimental constraints and show that such a notion, considered
cosmologically, can be subsumed in the dark-energy equation of
state in an expanding Friedman-Robertson-Walker (FRW) universe. 
Planned measurements of the temporal variation in redshift have the
potential to distinguish the possibilities. 
\end{abstract}

\maketitle

The OPERA collaboration has recently observed, with a 
significance of 6.0$\sigma$ if statistical and systematic
uncertainties are combined in quadrature, 
that neutrinos traverse a known length 
faster than they
would were the speed of light in vacuum assumed~\cite{opera}, 
signalling an apparent violation
of the theory of special relativity. 
In the face of this extraordinary outcome~\cite{reltest} alternate possibilities
must be considered. 
Let us set aside obvious and admittedly more probable 
%ultimately more plausible
 possibilities from the onset: we shall assume that 
the significance of the experimental result is robust and thus not 
obviated by unknown systematic errors, and that the OPERA collaboration has 
measured the group, rather than the phase, 
velocity of the ``muon neutrino'' wave packet. The empirical reality of 
neutrino oscillations~\cite{nuosc} 
complicates the simple picture of a propagating $\nu_\mu$; the propagating 
mass eigenstates are not those of flavor. Nevertheless, we 
assume that such details are not important here, so that
the apparent violation of special relativity is manifest. 
In this context, then, is it possible to reconcile the OPERA result with 
the theory of special relativity? An affirmative answer requires an adjustment
in our way of thinking. 
Consider that terrestrial measurements of the fundamental speed of light are made
under conditions in which only known matter is clearly absent, 
but cosmology tells us we 
live in a dark-energy and dark-matter dominated universe~\cite{concord}. 
I propose that neutrinos interact more weakly with the dark universe than
photons and all other known matter do, so that propagating neutrinos offer a better
measure of Einstein's limiting velocity. In this picture the photon remains 
massless, so that classical electrodynamics is unaltered, but interactions with the 
dark universe retard its speed slightly. This is tantamount to an index of refraction 
which differs slightly from unity, so that $c$ can be less than $c_\nu$, the neutrino speed. 
This possibility is not at odds with special relativity, for 
the photon speed is measured in a background of dark energy and dark matter, 
rather than in a vacuum devoid of such content. 
We know little about the nature of dark matter, and less about dark energy;
nevertheless, severe constraints do exist on the nature of the interactions
we posit with the dark sector. We consider them 
carefully in what follows before 
turning to cosmological tests of the picture we espouse. 

{\em Present Constraints.---} 
The empirical value of $c$ 
is $2.99792458\cdot 10^8\,\hbox{m/s}$~\cite{nist}. 
We employ it as a conversion factor to relate time to distance, and indeed use it to define the meter. 
Its precise numerical value, however, does not 
derive from any known fundamental principle, so that it in itself is not sacrosanct. 
Stringent experimental tests, both terrestrial and cosmological, speak to
the nature of the speed of light and hence to constraints on the interactions with
the dark universe we propose. Severe limits exist on the variation of 
the speed of light with frequency~\cite{schprl,AmelinoCamelia:1997gz}, as well as on 
variations in the speed of light with respect to the orientation of its velocity vector in 
space~\cite{hohenseeprl,hohenseeprd}. Moreover, its 
universality as a limiting value of speed is also well-established~\cite{greene,hohenseeprl}. 
Thus we suppose the needed interaction must be energy-independent, isotropic, and universal
for all matter save neutrinos. Under these conditions the speed of light, determined at the
present cosmological time, can remain the same in every inertial reference frame presently
accessible to us, albeit $c <  c_\nu$. 
%cosmic ray tests limited by lack of knowledge of the progenitor 

We assert that the speed of light $c$ and the neutrino limiting speed $c_\nu$ is related via
$c=c_\nu/n$, where $n$ is an index of refraction with $n>1$. 
In order to make our discussion concrete, we must employ an explicit 
framework for $n$. For definiteness suppose that we have a medium of scatterers, each of mass $M$, 
with mass density $\rho$. Employing the conventions and analysis of Ref.~\cite{svg_dcl} we have, for a 
photon of angular frequency $\omega$, 
\begin{equation}
n(\omega) = 1 + \frac{\rho}{4 M^2 \omega^2} {\cal M}_{\rm f} 
\end{equation} 
for $|n - 1|\ll 1$, and where ${\cal M}_{\rm f}$ is the forward Compton amplitude in the scatterer rest frame. 
Assuming the discrete symmetries parity, time-reversal, charge-conjugation, as well as
Lorentz invariance, unitarity, and analyticity, we have, for energies below particle-production
threshold, that ${\cal M}_f = \sum_{j=0} A_{2j} \omega^{2j}$ where $A_{2j} >0$ for $j\ge 1$~\cite{GGT}. 
To explain the OPERA data in the face of empirical constraints on the speed of light, we set
all $A_{2j}$ to zero save for $j=1$, so that $\rho A_2/4M^2$ is set by their result for 
$(v-c)/c$, namely, $\rho A_2/4M^2 = (2.48 \pm 0.28 ({\rm stat}) \pm 0.30 ({\rm sys}))\times 10^{-5} \equiv \delta_0$~\cite{opera}. 
We have posited a background of unknown matter, but our result, namely that the photon
sees an index of refraction which differs from unity, can be of broader origin. 
In particular, the index formula can be generalized to particles of zero mass through the 
introduction of a thermal bath~\cite{langliu,abbad}, and the structure of our expansion in $\omega$
remains unaltered. The stringency of the tests on nonobservation of anisotropies in the speed of light suggest that
our ``unknown matter'' is something other than the dark matter invoked to explain the observed galactic rotation curves. 
% neutrino-photon scattering really small!

The OPERA experiment is not the first to study arrival time differences of photons and 
neutrinos. A previous short baseline experiment was sensitive to deviations in 
$|v-c|/c$ to  $|v-c|/c < 4\times 10^{-5}$~\cite{shortbase}, and the MINOS collaboration has reported
a measurement of $|v-c|/c = 5.1 \pm 2.9 \times 10^{-5}$~\cite{minos}. The observation of
neutrinos from the supernova SN1987A~\cite{sn1987A} sets a much more stringent 
limit. The neutrinos were observed to arrive some 3 hours before the first detection of optical
brightening to yield a conservative limit 
%(and the absence of brightening had been confirmed to roughly an hour before that)
of $|v-c|/c \stackrel{<}{\sim} 2\times 10^{-9}$~\cite{longo}. 
The limit implicitly assumes that the 
initial neutrino and photon pulses were emitted simultaneously, though we do expect the thermal
neutrino burst from the core collapse, with neutrino energies of ${\cal O}(10\,\hbox{ MeV})$, 
to be emitted 
prior to the emission of visible light~\cite{nu_sn}. In Ref.~\cite{nu_sn} this time difference
is assessed at $\sim 10$ hours, for a red-supergiant progenitor, 
though taking the OPERA result at face value implies that 
the neutrino burst associated with SN1987A was emitted some 4 years after 
optical emission. 
Such a long time lag is puzzling, and perhaps even implausible, 
but it must be noted that the detailed mechanism of a core-collapse
supernova has not been established~\cite{burrows}. Moreover, 
SN 1987A in itself had many unusual features --- e.g., 
its progenitor star was  
a blue supergiant~\cite{arnett}. The observed luminosity 
was also roughly an order of magnitude smaller than a typical Type II supernova and may be the 
result of the denser makeup of the star~\cite{arnett}. 
Although the detected neutrino energies and burst duration 
appear consistent with their emission in 
core collapse leading to a proto-neutron-star~\cite{burrowslat}, 
the pulsar expected in this picture has not yet been observed. 
Alternate mechanisms are possible~\cite{wmac,supra,magnetar}, and features such as 
rotation of the core-collapse remnant and its associated magnetic fields 
may also play a role~\cite{magnetar}. 
In a black-hole--accretion-disk scenario, or ``collapsar'' model~\cite{wmac}, e.g., 
neutrinos are emitted 
from the edge of the accretion disk formed after the core collapse; they are also of 
MeV energy scale --- and the burst duration can be of comparable duration, though this 
outcome depends on the parameters of the model, as does whether the core-collapse 
neutrinos are trapped within the star~\cite{nagataki,mclsur}. 
Particular features of SN 1987A suggest it may have
had a companion star~\cite{1987Acomp} as well; perhaps 
the dynamics of a binary system help 
explain the needed time lag and burst duration, with the close association 
of the observed neutrino and optical bursts attributable to coincidence. 
Observations of neutrinos from gamma-ray bursts could well yield more discriminating 
limits~\cite{Jacob:2006gn}, but these have not yet been observed.

The need to confirm the OPERA result in an independent experiment is clear. 
Beyond this, the empirical determination 
of an energy dependence in the limiting speed of the neutrino 
would speak to complications beyond the simple picture we propose here. 
Recently constraints on superluminal models from 
the mere detection of neutrinos in the OPERA experiment have been 
discussed~\cite{CG,bi,cowsik}. These do not operate in the picture
espoused here because Lorentz symmetry is not broken at the level of particle interactions.
Pair bremsstrahlung, as discussed in Ref.~\cite{CG}, can nevertheless occur but via an explicit
interaction with the medium, much as computed in lepton ``trident'' production 
in quantum electrodynamics~\cite{brodskyting}. The observation of earth-crossing neutrinos in 
the OPERA energy range and beyond~\cite{earth} show that the presence of such 
pair production effects do not constrain our scenario. 
Confirmation of our particular scenario requires study of the speed of light in 
temporally different regimes. To realize this, we turn to cosmological studies. 

\begin{figure}
\includegraphics[width=5.0in]{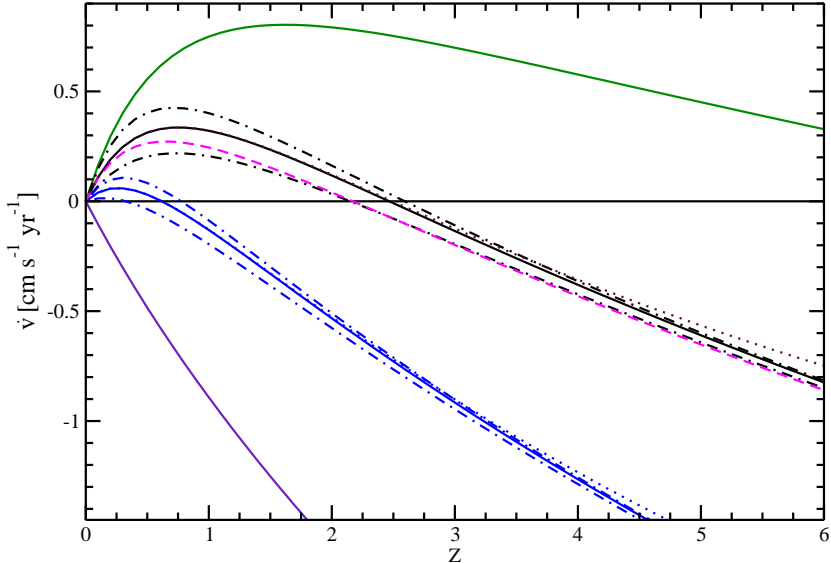}
\caption{The change in recession velocity $\dot{v}$ as a function of redshift $z$ in various, flat 
$\Lambda$CDM cosmologies, as well as in some alternate scenarios. 
The solid lines from bottom to top denote $\Omega_\Lambda=0,\, 0.5,\, 0.729$ (WMAP 7-year ``best fit''), and $0.9$, respectively, 
with $H_0=70.3$ km/s/Mpc\cite{Komatsu:2010fb}. 
The dot-dashed lines bracket the associated solid ones above and below 
when $w=-1$ is replaced by $w=-1.2$ and $w=-0.8$, respectively. 
The dashed curve results when $w$ is 
replaced by the z-dependent function used in Ref.~\cite{Carnero:2011pu} 
with $w_0=-1$ and $w_a=0.28$; the dotted curves result when 
the calculation of $\dot{v}$ is amended by an index of refraction as per the OPERA result.}
\label{fig1}
\end{figure}
{\em Time Variation in Redshift.---} The redshift to an object in a 
universe with matter and dark energy will change with time, and 
the measurement of its rate of change gives direct access to cosmological parameters~\cite{sandage,loeb,Corasaniti:2007bg}. 
The measurement is very challenging, and may well require decades of observing time~\cite{loeb,Corasaniti:2007bg}, 
%The change is very small, 
but with the advent of extremely large telescopes and laser-comb-stabilized calibration of spectrographs~\cite{murphy}, it
becomes possible~\cite{codex}. It is useful to recap the standard computation of ${\dot{z}}$~\cite{weinberg} before
turning to the inclusion of interactions. In an expanding FRW universe, the redshift $z$ of an object at time $t_1$ 
observed at time $t_0$ is related to the cosmological scale factor $a$ via 
$1+z = a(t_0)/a(t_1)$. Differentiating with respect to $t_0$ and noting $dt_1/dt_0 = 1/(1+z)$, we find~\cite{loeb} 
\begin{equation}
{\dot z} \equiv \frac{dz}{dt_0} = H_0(1+z) - H(z) \,,
\end{equation}
where the spectroscopic velocity shift is given by ${\dot v}= c{\dot z}/(1+z)$ and $H(z)$ 
is given in terms of 
\begin{equation}
H(z) = H_0 \left[ \Omega_M (1+z)^3 + \Omega_R (1+z)^4 + \Omega_{\Lambda} (1+ z)^{3(1+w)} + (1 - \Omega_{\rm tot})(1+z)^2
\right]^{1/2} \,,
\end{equation}
where $\Omega_M$, $\Omega_R$, and $\Omega_{\Lambda}$ represent the 
fraction of energy density in matter, radiation, and dark energy, respectively, relative to the critical 
density today. For the region of $z$ in which the stars and galaxies we observe reside, $\Omega_R$ is completely negligible, 
and, moreover, we shall assume a flat cosmology so that $\Omega_{\rm tot}=1$. 
The dark-energy density is characterized by an equation of state $w$, where $w=-1$ corresponds
to the cosmological constant. The value of $w$ need not be a constant in $z$, and in nonstandard cosmologies 
the scaling of the $\Omega_M$ term can also be modified, note Ref.~\cite{Corasaniti:2007bg} for a discussion. 
Curves illustrating the evolution of $\dot{v}$ with $z$ are shown in Fig.~\ref{fig1}. 
The possibility of light-dark-sector interactions can modify the shapes of these curves in $z$. 
The usual comoving distance is defined in the absence of interactions~\cite{dodelson}; to include
them via an index of refraction we note that the infinitesimal comoving distance is modified
from $\delta t/a$ to $n \delta t/a$ as the slowing of the speed of light makes the lightcone travel time 
longer~\cite{svg_dcl}. 
Consequently our starting point is modified to 
\begin{equation}
1+z = \frac{a(t_0)}{a(t_1)} \frac{n(t_1)}{n(t_0)} \,, 
\end{equation}
where we evaluate $n(z(t))$, noting in this case that 
$n(z)= 1 + (1+z)^3\delta_0$~\cite{svg_dcl}. 
Differentiating with respect to $t_0$ and noting that $dt_1/dt_0 = 1/(1+z)$ as before, as well as that
$dz(t_0)/dt_0=0$, yields 
\begin{equation}
{\dot z} = \frac{(1+z)H_0 - H(z)}{1- (1+z)\frac{d\ln n(z)}{dz}} \simeq \left((1+z)H_0 - H(z)\right)
\left(1 + (1+z)\frac{dn(z)}{dz}\right)
\end{equation}
to leading order in small quantities, where ${\dot v}= c_\nu{\dot z}/(1+z)$, though 
the replacement of the 
overall factor of $c$ with $c_\nu$ in $\dot{v}$ will always be insignificant as it appears in a product with $H_0$.
Note $|n-1| \sim {\cal O}(0.1)$ for $z\sim 10$. 
We compare the index-of-refraction-modified result for $\dot{v}$ with
that from the flat $\Lambda$CDM model, as well as with other scenarios, in Fig.~\ref{fig1}. 
The inclusion of $n$ shifts $\dot{v}$ to greater values with $z$, just as a value of $\Lambda>0$ itself does~\cite{svg_dcl}. 
We show, too, how $\dot{v}$ with $z$ changes if $w\ne -1$. 
%The manner in which $\dot{v}$ changes is 
Experimental constraints on $w$ exist largely for models with
constant $w$, so that, e.g., the WMAP 6-parameter $\Lambda$CDM fit to the 7-year data, 
combined with
data from baryon-acoustic oscillations (BAO) and 
the measured value of $H_0$, yields $w = - 1.10 \pm 0.14$~\cite{Komatsu:2010fb}. 
In comparison, 
a direct measurement of the BAO angular scale using a distribution of galaxies with $z= 0.5-0.6$ yields 
$w=-1.03\pm 0.16$ if the other parameters are fixed~\cite{Carnero:2011pu}. 
This data set also yields 
a constraint on $w(z)$: writing $w(z)= w_0 + w_a (1-1/(1+z))$ and 
using the WMAP 7-year ``best-fit'' parameters~\cite{Komatsu:2010fb}
yields $w_a=0.06\pm 0.22$~\cite{Carnero:2011pu}. 
Since the empirical data allow $w< -1$, we 
show how $\dot{v}$ changes if $w$ is altered to $w=-1.2$ or $w=-0.8$ in Fig.~\ref{fig1}, as well as if we employ
$w(z)$ and the WMAP 7-year ``best-fit'' parameters with $w_a=0.28$~\cite{Carnero:2011pu}. 
The effect of the modification with $n$ on $\dot{v}$ is rather small, though it begins to be appreciable
for $z$ in excess of $z\simeq 0.3-0.4$. Interestingly this is within the window of Lyman-$\alpha$ forest studies, 
for which peculiar motions are known to be negligibly small~\cite{rauch,codex}. 
If measurements of ${\cal O}(1\,\hbox{cm/s})$ can be made~\cite{murphy}, 
then a decade of observations can resolve differences
of ${\cal O}(0.1\,\hbox{cm/s})$ --- such a lengthy campaign is being planned~\cite{codex}.

{\em Conclusions.---} The sobering significance of the OPERA result~\cite{opera}, coupled with decades of successful tests 
of special relativity~\cite{reltest}, 
prompts us to consider alternatives which can be consistent with both. 
The experimental result $c < c_\nu$ need not reflect a breaking of Poincar\'e invariance but, rather, could 
speak to light-dark-sector interactions which yield an index of refraction which differs from unity. 
The interactions must be energy-independent and
isotropic and universal to all matter save neutrinos, to be consistent with existing experimental results. 
Nevertheless, the suggestion can be tested through the measured time-variation in redshift~\cite{sandage,loeb,Corasaniti:2007bg};
such studies permit the direct assessment of cosmological parameters. 
We have determined the amendment to $\dot{z}$ which appears in the 
presence of an index of refraction. To realize a simple but definite form of $n(z)$ we have 
asserted that the photon couples to unknown matter, yielding $n(z) \sim (1+z)^3$. Reality could be much
richer, and the possibilities include not only interactions with dark energy but also modifications to gravity itself. 
The interactions of which we speak may also be specific to very low redshift~\cite{Mortonson:2009qq}. 
Alternatively, perhaps $c_\nu > c$ can mediate additional radiative effects in the very early universe~\cite{Calabrese:2011hg}. 
Moreover, the notion
that a cosmologically local speed of light is tied to a dark energy model in 
which $w(z) +1 \ne 0$ is possible irrespective of whether OPERA is correct. 
The causal velocity itself could change with cosmological epoch, 
providing a context for the study of the time-dependence of other fundamental constants, 
such as $\alpha \equiv e^2/\hbar c$~\cite{webb}. 
Perhaps the OPERA result opens a new window on the dark universe --- and the
observational studies retain their interest even if it does not. 
The breadth of the possibilities 
underscores the importance of the observational measurement of 
both dark energy and its equation of state in a range of cosmological epochs. 

{\em Acknowledgments.---}
I gratefully acknowledge hospitality from the Aspen Center for Physics during a portion 
of this work, and I thank Shamit Kachru for comments on modified gravity models. 
I thank Scott Dodelson and Tim Gorringe for remarks helpful to improving the manuscript
and Stan Brodsky, Maury Goodman, and Gail McLaughlin for useful information. 
This work is supported, in part, by the U.S. Department of Energy under 
contract DE--FG02--96ER40989.

\end{document}